\documentclass{article}
\usepackage{lineno}
\usepackage{icrctc07}

\title{Constraints on top-down models for the origin of UHECRs from the Pierre Auger Observatory data}
\shorttitle{Constraints on top-down models}
\authors{D.V. Semikoz$^1$, for the Pierre Auger Collaboration$^2$}
\afiliations{$^1$ APC, 10, rue Alice Domon et Leonie Duquet, 75205 Paris, France \\
$^2$ Pierre Auger Observatory, av. San Martin Notre 304 (5613) Malarg\"ue, Argentina}
\email{dmitri.semikoz@apc.univ-paris7.fr}

\abstract{Taking into account the Pierre Auger Observatory limits on the photon fraction among the highest energy cosmic rays, we show that the models based on the decay of super-heavy dark matter in the halo of our Galaxy are essentially excluded from being the sources of UHECRs unless their contribution becomes significant only above $\sim 100$~EeV. Some top-down models based on topological defects are however  compatible with the current data and may be best constrained in the future by the high-energy neutrino flux limit.}
\begin{document}
\maketitle
\vspace{-12pt}
\section{Introduction}

Ultra-high-energy cosmic rays (UHECRs) raise a number of observational
as well as theoretical problems. While accelerating particles up to
energies above $10^{20}$~eV appears challenging in even the most
efficient astrophysical accelerators, a large variety of so-called
\emph{top-down models} have been proposed in which UHECRs are not
accelerated from ambient, low-energy particles, but produced directly
at the ultra-high energy from the decay of putative supermassive particles
with masses in excess of $\sim 10^{21}$~eV. The two main classes of such
models can be considered, with distinct generic properties. The first
one involves the decay or annihilation of topological defects (TDs)
produced through a phase transition in the early
universe~\cite{topdefects}. Such events would occur roughly
homogeneously throughout the universe, and generate supermassive
particles that would in turn decay into quarks and leptons and lead to
secondary UHECR protons and photons with an energy spectrum and
relative abundances characteristic of the underlying hadronization
process. These UHECRs would then propagate through the universe in
much the same way as if accelerated by astrophysical sources,
interacting with the CMB photons to produce e$^{+}$e$^{-}$ pairs and
pions (in the case of protons) -- the so-called \emph{GZK effect}~\cite{GZK1966}.

\begin{figure*}[ht]
\begin{center}
\includegraphics[width=0.3\textwidth,angle=270,clip]{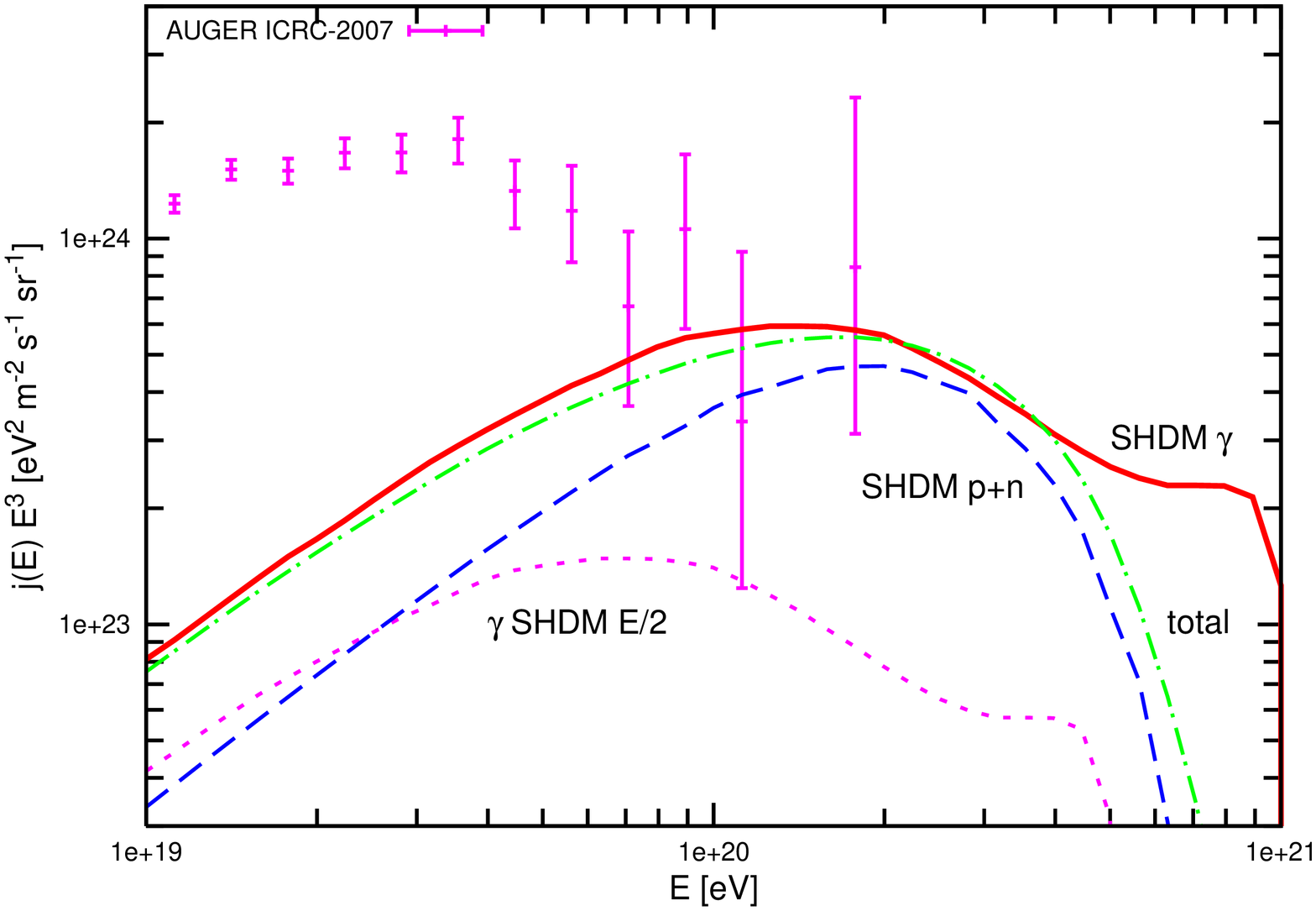}
\hfill\includegraphics[width=0.3\textwidth,angle=270,clip]{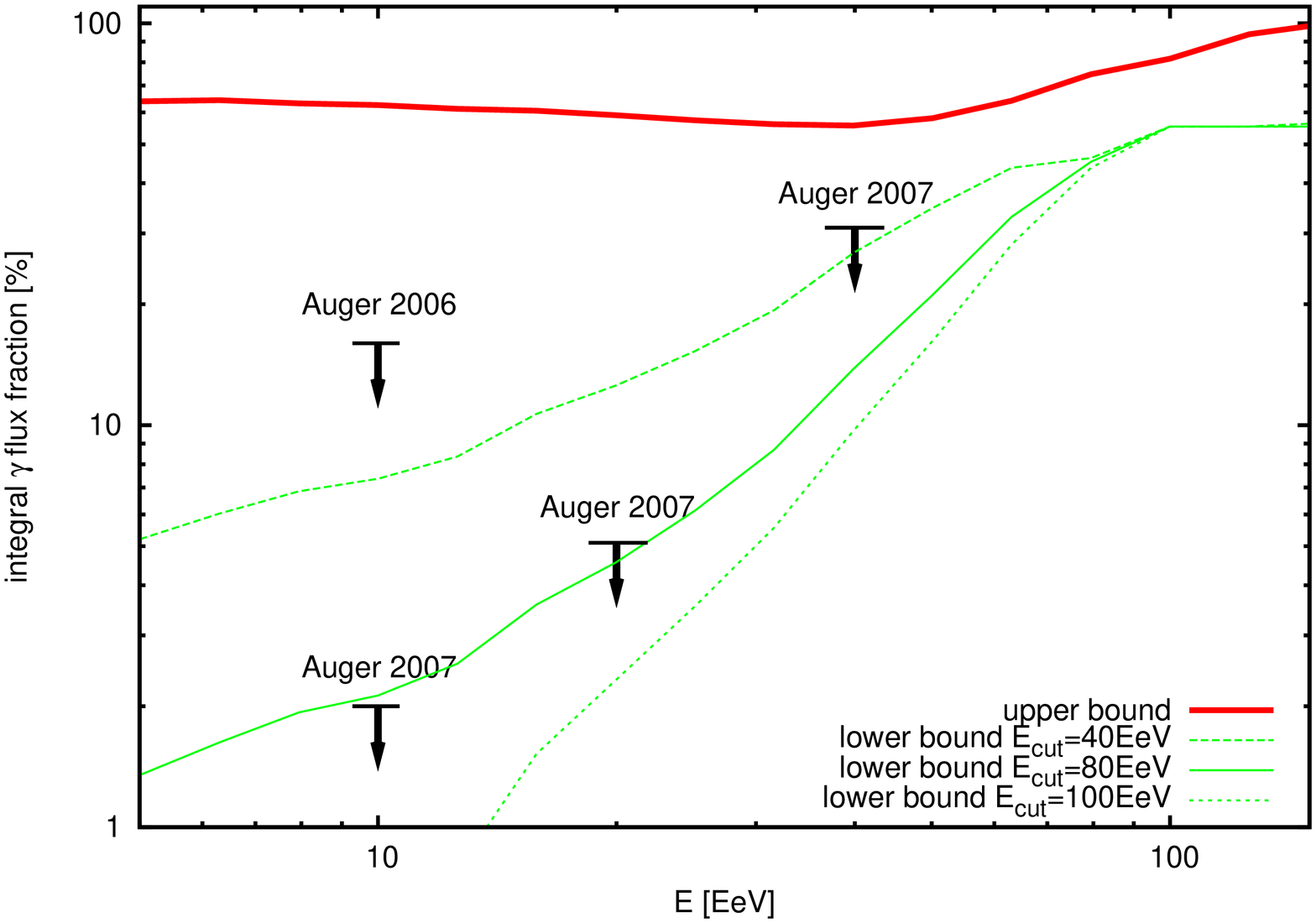}
\caption{\label{SHDMPhotonLimit} 
a (left): Example of SHDM decay products fit to the Pierre Auger Observatory spectrum. Solid line shows the photon flux from SHDM, the dashed line is the SHDM nucleon flux, the dotted line is photon flux reconstructed as protons (in approximation $E=E_\gamma/2$) and the dot-dashed line is the total reconstructed spectrum from SHDM. b (right): Corresponding photon fraction in percentage of the total UHECR spectrum integrated above energy $E$. Photon fraction located in the range between red and green lines with assumption that SHDM is responsible for the observed UHECR flux above 40 EeV, 80 EeV and 100 EeV.  ``Auger 2006''  and ``Auger 2007'' upper limits are from~\cite{AugerPhotons2006} and~\cite{AugerPhotonLimit}, respectively.}
\end{center}
\end{figure*}

In the second class of top-down models, the supermassive particles
responsible for the observed UHECRs are produced directly in the early
Universe~\cite{SHDM} and have a lifetime larger then the age of the Universe.
 An important motivation for such a scenario is that these
particles could make up the inferred dark matter in the universe,
which also provides a natural link between cosmology and UHECRs,
allowing one to relate their expected flux to the properties of
these super-heavy dark matter (SHDM) particles. A key feature of SHDM
scenarios is that the main contribution to the highest energy cosmic
rays observed at Earth would be provided by supermassive particles
concentrated in the halo of our Galaxy, so that propagation effects
(including the GZK suppression of the spectrum) would be
supressed by the two orders of magnitude~\cite{SHDM}. 
These models have thus been extensively studied in the
context of the report by the AGASA experiment of an excess of UHECR
events above $10^{20}$~eV~\cite{agasa}. However, neither the spectrum
reported by the HiRes experiment (suggesting the presence of the
expected GZK suppression~\cite{hires}) nor the Auger spectrum by
themselves can rule out SHDM models.

A detailed review of the various top-down models can be found in
ref.~\cite{bs}. 
A review of previous UHE photon~ limits can be found in 
ref.~\cite{photon_review} and a summary of neutrino~ limits in
ref.~\cite{neutrino_review}.
 In this paper, we analyse the constraints set on both
TD and SHDM models by the data of the Pierre Auger Observatory,
normalizing the UHECR flux to the Auger spectrum~\cite{AugerSpectrum}
and using the derived photon limit~\cite{AugerPhotonLimit} and
neutrino flux limit~\cite{AugerNeutrinoLimit}.

\section{Super-heavy dark matter scenarios (SHDM)}

Most SHDM models predict UHECR fluxes dominated by
the contribution of the Milky Way halo, so that propagation effects 
on both the spectrum
and the composition are insignificant for the commonly used radio background 
models~\cite{biermann}. 
UHECRs should thus be observed
as produced, with a relatively hard spectrum up to a fraction of the
initial mass of the SHDM progenitors and a dominant photon component
(here we use the recent results of \cite{Barbot-Drees}). The
latter prediction can be tested with Auger, thanks to the
photon/hadron discrimination power of both the surface detector and
the fluorescence detector~\cite{AugerPhotons2006,AugerPhotonLimit}. In
Fig.~\ref{SHDMPhotonLimit}a, we show a fit of the highest energy part of 
UHECR spectrum by
SHDM decay products, assumed to account for the measured flux above
 $8\,10^{19}$~eV. 
In addition to the primary photon flux, we show the
``apparent flux'' as would be reconstructed by the Auger analysis
procedure assuming proton primaries, through which a photon of energy 
$E_{\gamma}$ would be typically misinterpreted as a proton of energy 
$E_{\mathrm{p}} = E_{\gamma}/2$. This ``apparent'' photon component, 
added to the nucleon flux, makes up the total inferred UHECR spectrum.

\begin{figure*}[ht]
\begin{center}
\includegraphics[width=0.3\textwidth,angle=270,clip]{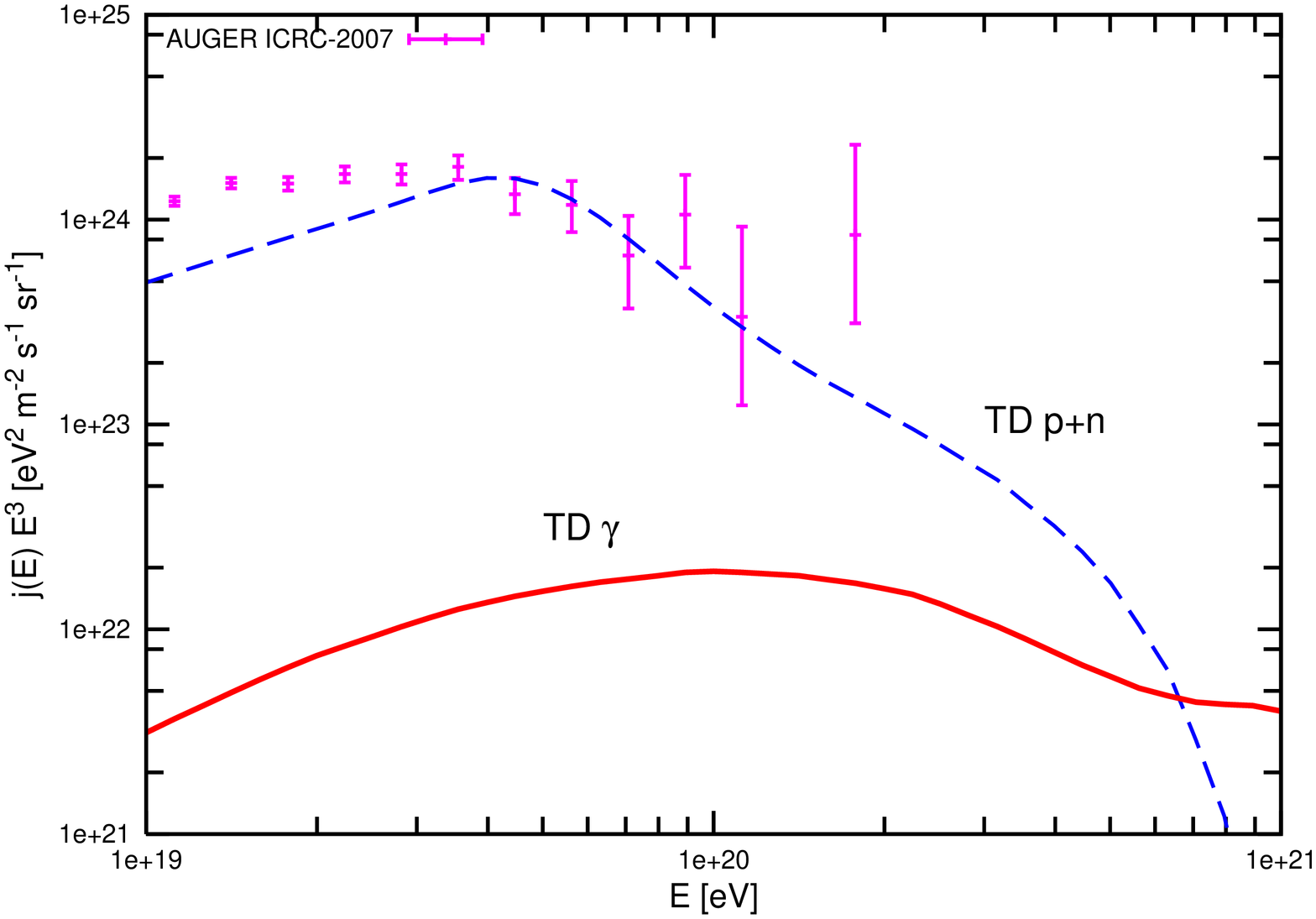}
\hfill\includegraphics[width=0.3\textwidth,angle=270,clip]{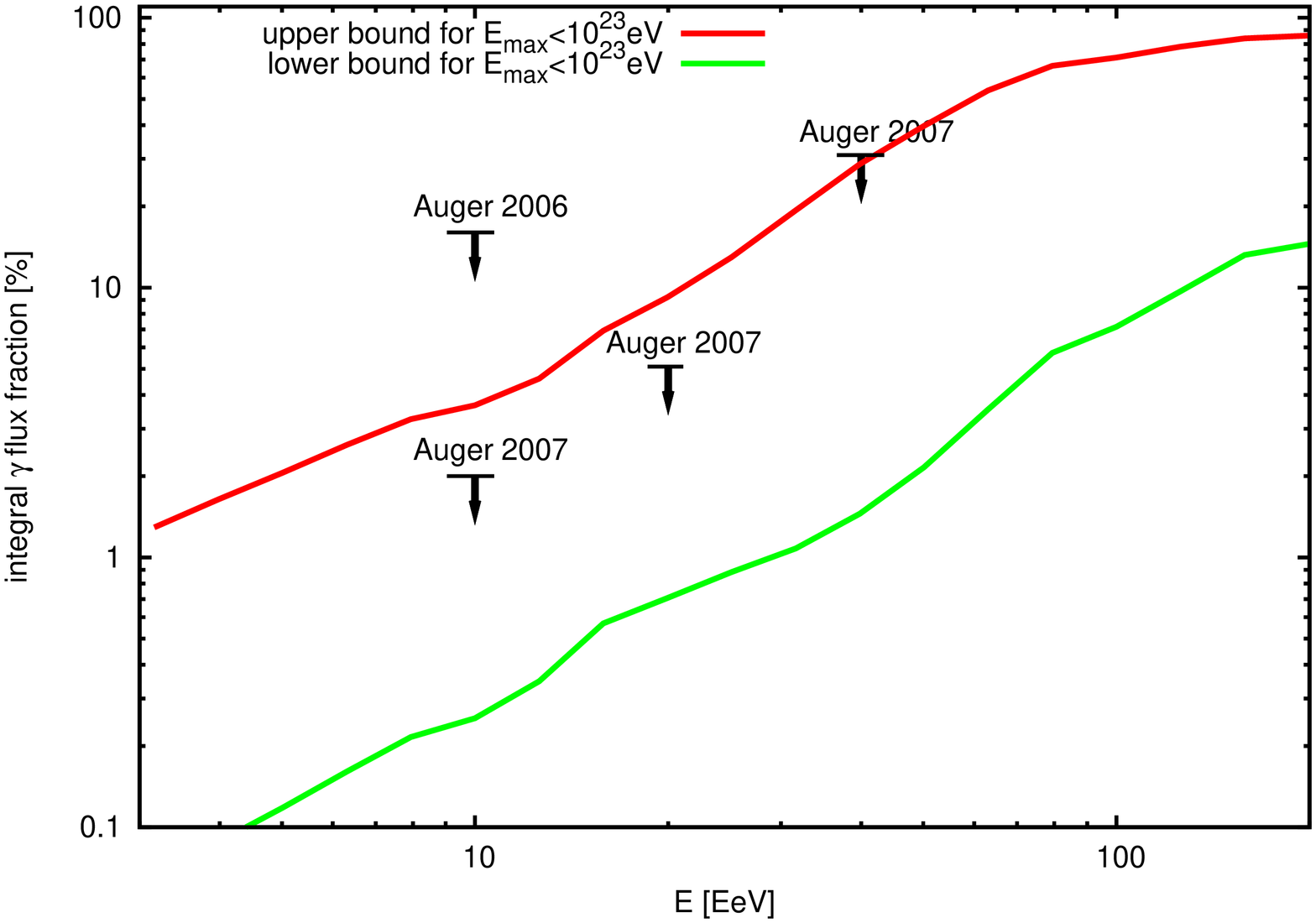}
\caption{\label{fig:TDSpectrum} a (left): Example of a fit of the Auger spectrum with nucleons (dahsed line) and photons (solid line) arising from TDs down to 40~EeV. b (right): Photon fractions among UHECR, as in Fig.~\ref{SHDMPhotonLimit}, for  super-heavy particle masses, $M_\mathrm{SH} = 2E_{\max}<2\times 10^{23}$ eV.}
\end{center}
\end{figure*}

To explore the parameter space of SHDM models, we fit the last few bins of the 
spectrum (above $4\,10^{19}$~eV,  $8\,10^{19}$~eV, or $10^{20}$~eV) using 
different values for the mass and normalization amplitude, according to the 
procedure described in~\cite{method}.

In Fig.~\ref{SHDMPhotonLimit}b, we show 
the corresponding photon fraction for two extreme cases of SHDM scenarios, 
giving the largest (upper curve) or lowest possible photon fraction (three lower 
curves). The latter are derived under the assumption that SHDM particles provide 
the dominant contribution to the UHECRs above the three indicated energies, 
respectively. The superimposed experimental limits show that SHDM top-down 
scenarios are ruled out by the Auger photon limit, except if they only 
contribute significantly to the cosmic-ray flux above $10^{20}$~eV. Therefore, 
SHDM models can only have a subdominant contribution to essentially all the 
UHECRs observed so far. 

\section{Scenarios involving Topological Defects (TDs)}

The injection spectrum and composition of UHECRs produced by the decay or 
annihilation of TDs are similar to those of the SHDM case. However, since TDs 
are expected to be evenly distributed throughout the universe, propagation 
effects are important and photons are strongly suppressed by their interactions 
with the extragalactic photon backgrounds. Their energy loss length is much 
smaller than that of protons up to above $10^{20}$~eV, so that the predicted 
photon-to-proton ratio \emph{at Earth} remains limited (much lower than the 
ratio at the source). In Fig.~\ref{fig:TDSpectrum}a, we show a fit of the 
highest energy part of the spectrum within a typical TD model, where the proton 
component dominates up to the highest detected energies. In 
Fig.~\ref{fig:TDSpectrum}b, the expected range of photon-to-proton ratios is 
shown for a variety of TD models (see, e.g., \cite{topdefects}), with different 
values of the maximum energies at the source. As can be seen, the current 
experimental limits only constrain the most photon-rich cases, and a wide range 
of models remain compatible with the data.  
Note that the model predictions are very
sensitive to the actual spectrum they are trying to account for. 
In particular, the results are different if one fits the
AGASA data or the HiRes data with SHDM and TD models~\cite{GZKphotons}.
In the case of the Pierre Auger Observatory it is more difficult to constrain
contribution of those models due to presence of the supression in the 
spectrum at highest energies~\cite{AugerSpectrum}.

\begin{figure}[ht]
\begin{center}
\includegraphics[width=0.3\textwidth,angle=270,clip]{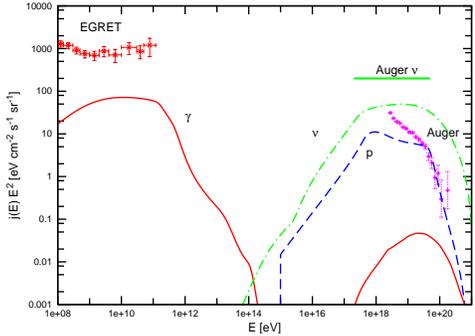}
\caption{\label{fig:CRsAndEGRET}   Photon (solid red line), neutrino (dashed-dotted green line), and proton (dashed blue line) fluxes from a TD model with cascade photons down to GeV. Also shown are Pierre Auger observatory UHECR spectrum~\cite{AugerSpectrum} and neutrino limits \cite{AugerNeutrinoLimit} and the EGRET~\cite{EGRET} limit on the diffuse gamma-ray background.}
\end{center}
\end{figure}

In Fig.~\ref{fig:CRsAndEGRET}, we plot the overall proton, photon and neutrino spectra 
associated with the propagation of UHECRs from a typical TD model, down to 100~MeV. 
This involves the secondary neutrinos produced by charged pion decay and the gamma-rays 
from the electromagnetic cascade induced by the UHE photons in the extragalactic medium. 
Such TD models are not constrained by the EGRET~\cite{EGRET} limit on the diffuse 
gamma-ray background. 
The neutrino upper limit set by Auger at high energy is also shown. Larger 
statistics expected in the future may lead to the strongest constraints on these 
models from the Auger data, or possibly the detection of their neutrino counterparts.
Top-down photons should also be detected eventually in this case.

\section{Discussion}

We have analysed the implications to top-down UHECR source scenarios of three 
complementary experimental results of the Pierre Auger Observatory presented in 
this conference: the energy spectrum, the photon fraction limit and the neutrino 
flux limit. We found that super-heavy dark matter models are strongly 
constrained by the absence of identified photon candidates in the Auger data. In 
particular, they cannot provide the dominant contribution to the overall UHECR 
flux at any energy below $\sim 10^{20}$~eV, which strongly restricts their 
motivation. On the other hand, models involving topological defects generally 
predict photon fractions after propagation that are compatible with the current 
data, while their neutrino counterpart may be more strongly constrained by 
future neutrino flux limits obtained with the Pierre Auger Observatory.

\end{document}